\documentclass[prl,showpacs,superscriptaddress,twocolumn,preprintnumbers,amsmath,amssymb,floatfix]{revtex4}
\usepackage{graphicx,natbib} 
\usepackage{times}
\usepackage{bm}
\usepackage{amssymb,amsmath}
\usepackage{epsfig}

\newcommand{\brokt}[3]{\left\langle #1 \right| #2 \left| #3\right\rangle} 
\newcommand{\focc}{$f^\mathrm{occ}~$ }
\newcommand{\funocc}{$f^\mathrm{un}~$}
\newcommand{\veck}{\mathbf{k}}
\newcommand{\eigvec}{\psi}      
\newcommand{\enk}{\epsilon_{n\veck}}  
\newcommand{\Enk}{\epsilon_{n\veck}^\mathrm{QP}}  

\newcommand{\dvu}{\delta \hat{V}^U}

\begin{document}
\title{The \textit{f}-electron challenge: localized and itinerant states in lanthanide oxides united by \textit{GW}@LDA+\textit{U}}
\author{Hong Jiang}
\affiliation{Fritz-Haber-Institut der Max-Planck-Gesellschaft, 14195 Berlin, Germany}
\author{Ricardo I. Gomez-Abal}
\affiliation{Fritz-Haber-Institut der Max-Planck-Gesellschaft, 14195  Berlin, Germany}
\author{Patrick Rinke}
\affiliation{Fritz-Haber-Institut der Max-Planck-Gesellschaft, 14195 Berlin, Germany}
\affiliation{University of California at Santa Barbara, CA 93106,USA}
\author{Matthias Scheffler}
\affiliation{Fritz-Haber-Institut der Max-Planck-Gesellschaft, 14195 Berlin, Germany}
\affiliation{University of California at Santa Barbara, CA 93106,USA}
\pacs{71.10.-w,71.15.-m,71.20.-b,71.27.+a}
\date{\today}
\begin{abstract}
Many-body perturbation theory in the $GW$ approach is applied to lanthanide oxides, using the local-density approximation plus a Hubbard $U$ correction (LDA+$U$) as the starting point. Good agreement between the $G_0W_0$ density of states and experimental spectra is observed for CeO$_2$ and Ce$_2$O$_3$. Unlike the LDA+$U$ method $G_0W_0$ exhibits only a weak dependence on $U$ in a physically meaningful range of $U$ values. 
For the whole lanthanide sesquioxide (Ln$_2$O$_3$) series $G_0W_0$@LDA+$U$reproduces the main features found for the optical experimental band gaps. The relative positions of the occupied and unoccupied $f$-states predicted by $G_0W_0$ confirm the experimental conjecture derived from phenomenological arguments.
\end{abstract}
\maketitle

The accurate first-principles description of the electronic structure of $f$-electron systems, i.e. materials containing lanthanide or actinide elements, is currently regarded as one of the great challenges in condensed matter physics. $f$-electron systems are characterized by the simultaneous presence of 
itinerant (delocalized) and highly localized $f$-states and interactions between them. Most computational methods are suited only to one type. Density-functional theory (DFT) -- currently the standard approach for electronic structure calculations of extended systems -- proves to be inadequate for $f$-electron systems in the most commonly applied local-density or generalized gradient approximation (LDA or GGA, respectively). One of the major deficiencies of LDA and GGA is the delocalization (or self-interaction) error \cite{Cohen08Sci}, which is particularly severe for systems with partially occupied $d$ or $f$-states and can even lead to qualitatively incorrect metallic ground states for many insulating systems. Hybrid functionals \cite{Becke93}, on the other hand, partly correct the self-interaction error by incorporating a certain portion of exact exchange, which significantly improves the descriptions of $d$- or $f$-electron systems \cite{DaSilva07a,Kudin02}. The dependence on adjustable parameters, however, remains a concern. Conversely, correlation effects that govern the physics of localized $f$-electrons can in principle be treated systematically by dynamical mean field theory (DMFT) \cite{Kotliar06RMP}. 
In practice these many-body corrections are only applied locally  to an atomic site (e.g. the Anderson impurity model) and the impurity solvers require input parameters (such as the Hubbard $U$) for the interaction strength. Moreover, most existing DMFT schemes are coupled (non self-consistently) to local or semilocal DFT calculations and the description of the itinerant electrons therefore remains on the level of LDA and GGA. 

As a first step towards a systematic \textit{ab initio} understanding of $f$-electron systems, we apply many-body perturbation theory (MBPT) in the $GW$ approach to a selected set of lanthanide oxides (CeO$_2$ and Ln$_2$O$_3$ (Ln=lanthanide series)) \cite{footnote:details} in this Letter. These compounds have important technological applications 
\cite{Lal88,Prokofiev96,Adachi98CR,Petit05,Singh06,REOBook2007}, in particular in
catalysis, where CeO$_2$-based compounds have attracted considerable interest from both
experiment and theory
\cite{Trovarelli02CeriaBook,Wuilloud84,Marabelli87,Mullins98,Skorodumova01,DaSilva07a,Castleton07,Fabris05R,Pourovskii07}.
Unlike in most previous studies, the $GW$ calculations in this work are based on LDA ground state calculations including a Hubbard $U$ correction (henceforth denoted $G_0W_0$@LDA+$U$).   
Our $G_0W_0$@LDA+$U$ calculations provide a qualitative understanding of the general trend observed for the band gaps of the Ln$_2$O$_3$ series and reproduce the characteristic features of the series, in particular the four dips observed in the experimental curve.

The $GW$ approach corresponds to the first order term of a
systematic expansion in MBPT  \cite{Hedin65} and has become the method of choice for the description of quasiparticle band structures in weakly correlated solids
\cite{Aryasetiawan98RPP_Onida02RMP}. Through the screened Coulomb interaction $W$ it captures the screening among itinerant electrons while at the same time treating exchange at the exact exchange level (given by the Hartree-Fock expression). 
The latter should
account for a large part of the many-body interactions among localized $d$ or $f$-electrons, as demonstrated recently for Cu$_2$O and VO$_2$ \cite{Gatti07bBruneval06b}. In this Letter we challenge the conventional view that regards many $f$-electron systems as
\emph{strongly correlated} electron systems for which band theory is
inadequate \cite{Pantelides07}. In the hierarchy of many-body perturbation theory,
strong correlation denotes correlation effects that go beyond exact
exchange and the weak correlation regime of $GW$. 
The good agreement between our $GW$ calculations and available experimental data
demonstrates that the $GW$ method can treat both itinerant $spd$
bands and localized $f$-bands accurately for the materials we have
considered. 
Since the screened exchange picture of the $GW$ approach captures the essential physics we challenge the classification of these materials as  \emph{strongly correlated}.

\begin{figure*} 
   \epsfig{bbllx=48,bblly=53,bburx=591,bbury=252,
           file=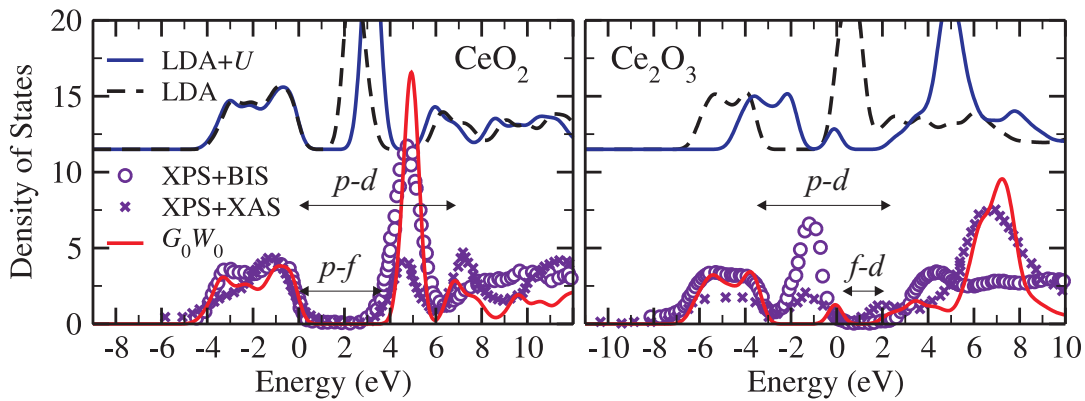,width=0.80\linewidth,clip}
\caption{\label{fig:dos} (Color online) The DOS of CeO$_2$ (left) and Ce$_2$O$_3$ (right) from LDA, LDA+$U$ and $G_0W_0$@LDA+$U$ with $U$=5.4 eV are compared to experimental data(XPS+BIS and XPS+XAS) \cite{footnote:fig-dos}. The LDA and LDA+$U$ curves have been offset vertically for clarity. 
}
\end{figure*}

The $GW$ method is typically applied in a perturbative manner (henceforth denoted 
$G_0W_0$) in which the quasiparticle (QP) energies $\Enk$ are calculated as a first-order correction to the eigenenergies $\enk$ and eigenvectors $\eigvec_{n\veck}$ of a reference single particle Hamiltonian $\hat{H}_0$
\begin{equation}
\label{eq:qp}
 \Enk=\enk+\Re\brokt{\eigvec_{n\veck}}{\Sigma^{\rm xc}(\Enk) - V^{\rm xc}} {\eigvec_{n\veck}}
\end{equation}
Here $\Sigma^{\rm xc}$ is the $G_0W_0$ self-energy calculated from the one-particle Green's
function $G_0$ and screened Coulomb interaction $W_0$, both evaluated using
$\enk$ and $\eigvec_{n\veck}$, and $V^{\rm xc}$ is the exchange-correlation potential included in $\hat{H}_0$. For most of the sesquioxides considered in this work LDA and GGA incorrectly predict a metallic ground state. In these cases first order perturbation theory based on LDA or GGA is not applicable and alternative reference $\hat{H}_0$ have to be employed \cite{Rinke08PSSRinke05,vanSchilfgaarde06b,Bruneval06,Shishkin07b}. 
In this work we use the LDA+$U$ method \cite{Anisimov93_97JPCM} as the starting point for $G_0W_0$. By adding a site- and orbital-dependent correction $\dvu$ to the standard LDA single-particle Hamiltonian, LDA+$U$ significantly improves the description of highly localized states, and therefore overcomes the major failure of LDA for these systems. To describe highly localized states accurately, we have implemented an all-electron $G_0W_0$ approach \cite{GomezAbal08} based on the full-potential linearized augmented plane wave method \cite{Wien2k}.

The LDA+$U$ method is conceptually similar to $GW$. It is, however, \textit{not} a substitute for $GW$ even for localized states: 1) The link between the LDA+$U$ and the $GW$ approximation relies on the assumption that the hybridization between localized and itinerant states can be neglected, which in many cases is not valid; 2) the $\dvu$ correction in LDA+$U$ has \emph{direct} effects only on the corresponding localized states; the description of itinerant states remains at the LDA level;  and 3) screening in LDA+$U$ is \textit{static}, while in reality screening is dynamic and, has a stronger energy-dependence for localized electrons than for itinerant ones. The LDA+$U$ approach by itself is therefore not expected to provide a satisfactory description to the electronic structure of $f$-electron systems. 

An advantage of the $G_0W_0$@LDA+$U$ approach lies in the fact that the Hubbard $U$ corrections enter self-consistently in the ground state calculation. This becomes important when localized states hybridize with band states.
A less appealing aspect of the LDA+$U$ approach concerns the parameter $U$, which, in many cases, is determined by fitting to experimental data. 
The onsite Coulomb interaction $U$, however, has a well-defined physical meaning, and can be calculated
from first-principles (see, e.g. Refs. \cite{Anisimov91aCococcioni05}).
In addition we demonstrate below that $G_0W_0$
based on LDA+$U$ is much less sensitive to $U$ than the LDA+$U$ itself. The $G_0W_0$ calculations also remove the problem of the double counting corrections that are not well defined in the LDA+$U$ approach, by subtracting them out as part of $V_{xc}$ in  Eq. \ref{eq:qp}.
Despite these advantages the $G_0W_0$@LDA+$U$ approach will not be suitable in cases where strong correlation effects become important (e.g. the Kondo resonance), for which many-body interactions that go beyond the $GW$ approach have to be included. This is currently the domain of DMFT, as alluded to in the introduction.

Figure \ref{fig:dos} shows the density of states (DOS) of CeO$_2$ and Ce$_2$O$_3$
calculated from LDA, LDA+$U$, and $G_0W_0$@LDA+$U$ (with $U$ =  5.4~eV \cite{footnote:U}) together with the experimental spectra. The $G_0W_0$ density of states for CeO$_2$ agrees well with the experimental data from direct (XPS) and inverse  (BIS) photoemission spectroscopy or X-ray absorption spectroscopy (XAS). In CeO$_2$, the empty $f$-states introduce a sharp peak in the fundamental band gap formed between the O-2$p$ valence and Ce-5$d$ conduction band. The most important quantities here are therefore the $p$-$f$ and $p$-$d$ gaps. For the latter the $G_0W_0$ value of 6.1~eV is in good agreement with the experimental one of 6.0~eV. The  $p$-$f$ gap, however, cannot be unambiguously determined from XPS-BIS or other available measurements (See, e.g. Ref. \cite{Castleton07} and references therein).  As expected, LDA underestimates both gaps, but the $p$-$d$ gap is only slightly smaller than in experiment (5.5 vs 6.0 eV), whereas LDA+$U$ decreases it to 5.1~eV.
We also note that our $G_0W_0$@LDA+$U$ DOS for CeO$_2$ is very close to that obtained from the recently proposed self-consistent $GW$ method \cite{vanSchilfgaarde06a}. 

More intriguing features are observed for Ce$_2$O$_3$. As expected, the on-site
Hubbard correction in the LDA+$U$ splits the single $f$-peak in LDA to occupied and unoccupied $f$-bands (denoted as \focc and \funocc, respectively), the former falling within the $p$-$d$ gap and the latter overlapping with the conduction bands. The $p$-$d$ gap is nearly the same as in LDA, but the $p$-\focc splitting is greatly reduced. 
Applying the $G_0W_0$ corrections to the LDA+$U$ ground state, we observe two remarkable features: 
1) the O-2$p$ band is pushed to lower energy with respect to the \focc band, and
2) the \funocc-band shifts up in energy away from the conduction band edge increasing the splitting between the \focc and \funocc bands at the same time.

The $G_0W_0$ and the experimental spectrum in Fig. \ref{fig:dos} are aligned at the upper valence band edge and not the Fermi level since this is not well defined at the 0 K at which the calculations are performed.  We again find the \funocc peak to be in good agreement with the BIS data. The position of the \focc peak, however, differs by approximately 1~eV. With respect to the band gap the comparison with experiment is aggravated by the limited experimental resolution. Taking the difference between the conduction band edge and the upper edge of the XPS-XAS \focc peak gives a band gap that is
consistent with the optical band gap of $\sim$2.4~eV \cite{Prokofiev96,footnote:optgap} and the $G_0W_0$ gap of 2.0~eV.  If one instead references against the peak center of the \focc states, the experimental band gap of Ce$_2$O$_3$ would be larger than 3 eV. Further experimental evidence is clearly needed to determine the actual value.
\begin{figure}
   \epsfig{bbllx=10,bblly=0,bburx=740,bbury=460,
           file=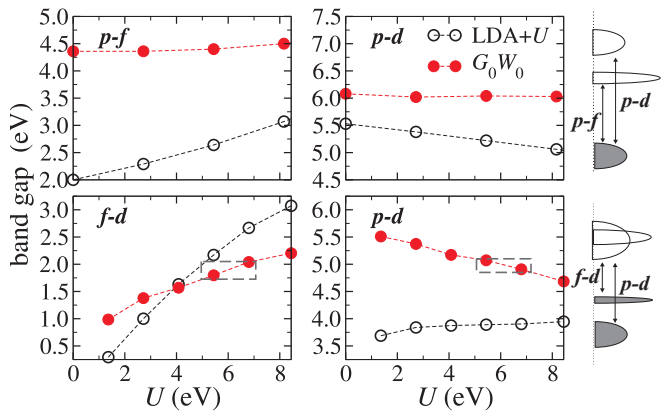,width=\columnwidth,clip}
\caption{\label{fig:Egap-U} (Color online) Band gaps of CeO$_2$ ($p$-$f$ and $p$-$d$, upper row)
and Ce$_2$O$_3$ ($f$-$d$ and $p$-$d$, lower row) from LDA+$U$ and $G_0W_0$@LDA+$U$ as a function of $U$. The dashed rectangles in the lower panel indicate the range of physically meaningful values of $U$ (see text). }
\end{figure}

Figure \ref{fig:Egap-U} illustrates the influence of $U$ on the LDA+$U$ and $G_0W_0$@LDA+$U$ calculations for the examples of CeO$_2$ and Ce$_2$O$_3$. Since the $f$-states are essentially empty in CeO$_2$, the effect of $U$ is relatively weak: For $U$=0 to 8 eV, the $p$-$f$ and $p$-$d$ gaps from LDA+$U$ change only by approximately 1.0 and 0.5 eV, respectively, and become nearly constant in $G_0W_0$. The situation is more complex in Ce$_2$O$_3$. In LDA+$U$, the $f$-$d$ gap depends sensitively on $U$ and varies by nearly 3 eV, but the $p$-$d$ gap remains almost unaffected. In contrast, both the $f$-$d$ and the $p$-$d$ gap exhibit a slight $U$-dependence in $G_0W_0$, changing by 1.2 and 0.6 eV over the full $U$ range explored here. Most importantly, however, the $U$-dependence reduces to only approximately 0.3~eV in the range of ``physically meaningful" values of $U$ ($\sim$5-7~eV \cite{Castleton07,Fabris05R}), which is already in the range of experimental error bars. We note in passing that the relatively weak dependence on $U$ has also recently been reported by Kioupakis \textit{et al.}, who applied a similar approach to solid hydrogen \cite{Kioupakis08}. In addition, Figure \ref{fig:Egap-U} shows that within LDA+$U$ one could obtain an apparently more accurate $p$-$f$ (in CeO$_2$) and $f$-$d$ (in Ce$_2$O$_3$) gap by using a significantly larger $U$, which, however, would not improve the description of itinerant states including, in particular, the $p$-$d$ gap. 

\begin{figure}
\centerline{\includegraphics[width=3.3in,clip]{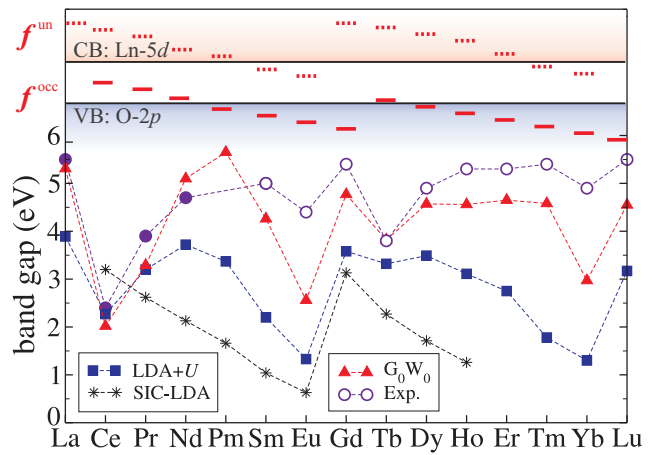}}
\caption{\label{fig:Egap-R2O3} (Color online) Band gaps of the Ln$_2$O$_3$ series from LDA+$U$ and $G_0W_0$ ($U$=5.4 eV) are compared to SIC-LDA results \cite{Petit05} and experimental optical gaps \cite{Prokofiev96}. The schematic in the upper part of the figure illustrates the position of the \focc and \funocc states extracted from the $G_0W_0$@LDA+$U$ calculations in relation to the valence and conduction band edge (VB and CB).}
\end{figure}

It has long been recognized that, although many properties of rare-earth compounds exhibit a monotonous behavior across the lanthanide series, some show a striking variation. For example the optical band gaps of rare earth sesquioxides \cite{Prokofiev96} shown in Fig.~\ref{fig:Egap-R2O3} exhibit clear dips  for Ce, Eu, Tb, and Yb, which appear to be unaffected by structural variation across the series. 
In Fig.~\ref{fig:Egap-R2O3}  the optical gaps of the Ln$_2$O$_3$ series \cite{footnote:optgap} are  compared to LDA+$U$, $G_0W_0$@LDA+$U$ and previous self-interaction corrected LDA (SIC-LDA) results \cite{Petit05}. Since $U$ is the effective e-e interaction among $f$-electrons \textit{screened by $spd$ states}, which are very similar in all Ln$_2$O$_3$ compounds, we expect $U$ to be only weakly dependent on the number of localized $f$-electrons, and therefore use a constant $U$=5.4~eV for the whole series. 
Compounds denoted by filled circles in Fig.~\ref{fig:Egap-R2O3} crystallize preferentially in the hexagonal structure, for which all calculations have been performed. Starting from Sm$_2$O$_3$ (denoted by open circles),  the most stable phase at room temperature is  cubic bixbyite, but the middle members can also exist in the monoclinic phase \cite{Adachi98CR}. 

As can be seen from Fig.~\ref{fig:Egap-R2O3} all the essential features in the experimental curve are reproduced by the $G_0W_0$  calculations including the four dips and the behavior in between. Even the quantitative agreement is good for most compounds. In addition our first principles calculations provide easy access to the character of each peak in the DOS and thus the character of the band gap, which is schematically shown in the upper part of Fig.~\ref{fig:Egap-R2O3}. 
In La$_2$O$_3$ (empty $f$-shell) the band gap is formed between the O-2$p$ valence and the La-5$d$ conduction band. As the occupation of the $f$-states increases, both \focc and \funocc continuously move downward in energy and the band gap evolves from $p$-$d$ via $f$-$d$ to $p$-$f$. This process repeats itself in the second part of the series (starting from Gd$_2$O$_3$) where the spin-up $f$-states have become fully occupied (and lie deep below the O-2$p$ states) and the spin-down $f$-states move downward in energy with increasing occupation. 

The character of the band gap across the series agrees well with the experimental conjecture derived from phenomenological arguments \cite{Lal88,Prokofiev96,vanderKolk06}. This is not the case in SIC-LDA and LDA+$U$. Not only does LDA+$U$ underestimate the band gaps of most Ln$_2$O$_3$ compounds, it also only shows a weak minimum at Tb$_2$O$_3$ and fails to reproduce the plateau between Ho$_2$O$_3$, Er$_2$O$_3$, and Tm$_2$O$_3$. 


The remaining quantitative differences between the $G_0W_0$ and experimental curve (especially for the later members of the series) could be due to several factors:  1) the experimental error bar, sample quality and instrumental resolution are not known \cite{Prokofiev96}; 2) to exclude the influence of excitonic effects the $G_0W_0$ results should be compared to photoemission (both direct and inverse) spectra, which are unfortunately not available for most members of the Ln$_2$O$_3$ series; 3) spin-orbit and multiplet  effects are not taken into account in our calculations; and 4) our $G_0W_0$ calculations are performed for the hexagonal structure, but the later members of the Ln$_2$O$_3$ series (after Ln=Sm) crystallize in the monoclinic or cubic bixbyite structure \cite{footnote:details}. Further investigations of these issues will be addressed in the future. 

We close this Letter with some remarks about strong correlation. 
A system is often commonly (but by no means satisfactorily) classified as strongly correlated if DMFT provides an improved description over LDA and it is therefore interesting to compare $G_0W_0$@LDA+$U$ and LDA+DMFT \cite{Kotliar06RMP}. 
Both methods add dynamic (i.e. energy dependent) effects to the LDA+$U$ Hamiltonian. LDA+DMFT is based on an underlying picture of \textit{localized states}, in which higher order correlation effects among the localized states can be easily incorporated. However, delocalized states (e.g. $spd$ states in $f$-electron systems) are still treated at the LDA level.
In contrast, $G_0W_0$ based on LDA+$U$ improves upon LDA+$U$ in the band picture and both localized and itinerant states are treated at the same level (i.e. $G_0W_0$). For Ce$_2$O$_3$ this can be directly quantified by comparing the $G_0W_0$@LDA+$U$ density of states with that from recent LDA+DMFT calculations by Pourovskii \textit{et. al.} \cite{Pourovskii07}. The $f$-states are described in a similar manner, but the $p$-$d$ gap in LDA+DMFT amounts to only $\sim$3.5 eV, which is significantly smaller than the experimental ($\sim$ 5.5 eV) and the $G_0W_0$@LDA+$U$ ($\sim$ 5.1 eV) value. For many open $d$- or $f$-shell compounds it can therefore be expected that the $GW$ approach can provide a consistent and accurate treatment for both localized and itinerant states and it will be illuminating to see when many-body effects beyond the $G_0W_0$ approximation become significant. 

We acknowledge fruitful discussions with Xinzheng Li, Xinguo Ren and Kris Delaney,
and the Nanoquanta Network of Excellence (NMP4-CT-2004-500198) for financial
support.


\begin{thebibliography}{40}
\bibitem[{foo({\natexlab{b}})}]{footnote:details}
\bibinfo{note}{We consider CeO$_2$ in the cubic fluorite structure, and Ln$_2$O$_3$ in the hexagonal (A-type) structure with antiferromagnetic ordering. Experimental lattice constants are used for CeO$_2$ and the light oxides (Ln$_2$O$_3$, Ln=La-Nd); for the heavy lanthanides (after Ln=Pm), we use lattice constants obtained by extrapolation from those of the light Ln$_2$O$_3$.}
\bibitem[{foo({\natexlab{a}})}]{footnote:fig-dos}
\bibinfo{note}{The XPS+BIS data for CeO$_2$ and Ce$_2$O$_3$ has been extracted
  from Refs. \cite{Wuilloud84} and \cite{Allen85}, respectively, and the
  XPS+XAS data from Ref. \cite{Mullins98}. The Fermi energy (the valence band
  maximum for insulating states) is taken as zero. The small peak at $\sim$2~eV
  in the BIS curve of Ce$_2$O$_3$ is probably due to the coexistence of the
  CeO$_2$ phase, as indicated in Fig. 3 of Ref. \cite{Allen85}}.
\bibitem[{foo({\natexlab{a}})}]{footnote:optgap}
\bibinfo{note}{Rigorously speaking the $GW$ band gap should be compared to the fundamental gap measured by   photoemission (both direct and inverse). The optical gap differs from the fundamental gap by electron-hole interaction effects  \cite{Aryasetiawan98RPP_Onida02RMP} At present the size of these so called excitonic effects is not known for lanthanide oxides.}
\end{thebibliography}
\end{document}